\documentclass[abstract=on,notitlepage,superscriptaddress,nofootinbib,aps,showpacs,twocolumn,prd]{revtex4}
\usepackage[latin1]{inputenc}
\usepackage{graphicx}
\usepackage{amssymb}
\usepackage{color}
\usepackage{float}
\usepackage{amsmath}
\usepackage{amsfonts}
\usepackage{dcolumn}
\usepackage{hyperref}
\usepackage{amsthm}
\usepackage{color}

\def\3nab{\tilde{\nabla}}

\def\be {\begin{equation}}
\def\ee {\end{equation}}
\def\ba {\begin{eqnarray}}
\def\ea {\end{eqnarray}}

\newtheorem{prop}{Proposition}
\newtheorem{cor}{Corollary}

\newcommand{\barray}{\begin{array}}
\newcommand{\earray}{\end{array}}

\newcommand{\bse}{\begin{subequations}} \newcommand{\ese}{\end{subequations}}
 
\begin{document}
\title{Conformal symmetries in generalised Vaidya spacetimes}
\author{Samson Ojako}
\email{ojakosamson@gmail.com}
\affiliation{Astrophysics and Cosmology Research Unit, School of Mathematics, Statistics and Computer Science, University of KwaZulu-Natal, Private Bag X54001, Durban 4000, South Africa.}
 \author{Rituparno Goswami}
\email{Goswami@ukzn.ac.za}
\affiliation{Astrophysics and Cosmology Research Unit, School of Mathematics, Statistics and Computer Science, University of KwaZulu-Natal, Private Bag X54001, Durban 4000, South Africa.}
\author{Sunil D. Maharaj}
\email{Maharaj@ukzn.ac.za}
\affiliation{Astrophysics and Cosmology Research Unit, School of Mathematics, Statistics and Computer Science, University of KwaZulu-Natal, Private Bag X54001, Durban 4000, South Africa.}
 \author{Rivendra Narain}
 \email{Narain@ukzn.ac.za}
\affiliation{Astrophysics and Cosmology Research Unit, School of Mathematics, Statistics and Computer Science, University of KwaZulu-Natal, Private Bag X54001, Durban 4000, South Africa.}

\begin{abstract}
In this paper we excavate, for the first time, the most general class of conformal Killing vectors, that lies in the two dimensional subspace described by the null and radial co-ordinates, that are admitted by the generalised Vaidya geometry. Subsequently we find the most general class of generalised Vaidya mass functions that give rise to such conformal symmetry. From our analysis it is clear that why some well known subclasses of generalised Vaidya geometry, like pure Vaidya or charged Vaidya solutions, admit only homothetic Killing vectors but no proper conformal Killing vectors with non constant conformal factors. We also study the gravitational collapse of generalised Vaidya spacetimes that posses proper conformal symmetry to show that if the central singularity is naked then in the vicinity of the central singularity the spacetime becomes almost self similar. This study definitely sheds new light on the geometrical properties of generalised Vaidya spacetimes. 
\end{abstract}
 
\maketitle

\section{Introduction\label{Intro}}

Symmetries provide us with a deeper insight into properties of the spacetime manifold and assist in finding new solutions to the Einstein field equations. Conformal symmetries are connected to the causal structure of spacetimes. If a spacetime admits conformal symmetry then there exists a conformal Killing vector field in the spacetime. If the metric is Lie dragged along this vector field the causal structure of the spacetime remains invariant. It is thus obvious that homothetic symmetry or Killing symmetries are special case of conformal symmetry. 
For general results connecting conformal symmetries to  kinematical and dynamical quantities see the investigations for anisotropic relativistic fluids by Maartens et al \cite{MaartensREllis1995}, Mason and Maartens \cite{MasonMaartens1985} and Coley and Tupper \cite{ColeyTupper}. Several examples of conformal Killing vectors have been found mainly in spherically symmetric spacetimes. In a recent study, Moopanar and Maharaj \cite{MoopMah2013,Israel_1967} obtained the general conformal Killing vector in spherically symmetric spacetimes. It is interesting to observe that homothetic and conformal symmetries arise in general {\it Locally Rotationally Symmetric} (LRS) spacetimes as shown by Singh et al \cite{EllisGF2017, SSGRSDM2019} using covariant semitetrad formalisms. It is known that the Vaidya spacetime admits a homothetic vector associated with a self-similarity. However, it does not admit a proper conformal Killing vector since it contains pure radiation as shown in Lewandowski \cite{lewandowski1990conformal}. The existence of proper conformal symmetries in generalised Vaidya spacetimes remains an open question as the  matter content is not only pure radiation and the conditions of Lewandowski are not applicable. This paper attempts to answer the question of existence of conformal symmetries in generalised Vaidya spacetimes.\\

The Vaidya spacetime \cite{Vaidya} is widely used in many astrophysical applications with strong gravitational fields. In general relativity, this spacetime assumed added importance with the completion of the junction conditions at the surface of the star by Santos \cite{santos1985non}. The pressure at the surface is nonzero and the star dissipates energy in the form of heat flux. This made it possible to study dissipation and physical features associated with gravitational collapse as shown by Herrera et al  \cite{herrera2006some, herrera2012dynamical}. Some recent studies of the temperature properties inside the radiating star include Reddy et al \cite{reddy2015impact}, Thirukkanesh et al \cite{thirukkanesh2012final} and Thirukkanesh and Govender \cite{thirukkanesh2013}. The metric in \cite{Vaidya, Lindquist_1965} may be extended to include both null dust and null string fluids leading to the generalised Vaidya spacetime. The properties of the generalised Vaidya metric have been studied by Hussain \cite{husain1996exact}, Wang and Wu \cite{Generalizedvaidya1999},  Glass and Krisch \cite{Radiationstring1998, twofluidatm1999}. Maharaj et al \cite{Radiatingstars2012} modelled a radiating star with a generalised Vaidya atmosphere in general relativity. Detailed study of continual gravitational collapse of these spacetimes in the context of the Cosmic Censorship Conjecture were done in \cite{Maombi1, Maombi2}. In the geometrical context, gravitational collapse has been considered in Lovelock gravity theory \cite{Gravitationalcollapse2011}, black holes in dynamical cosmology backgrounds \cite{SurroundedVaidya2018} and in electromagnetic fluids \cite{Heydarzvaidya2018}. The influence of dust, radiation, quintessence and the cosmological constant are included in these studies.\\

Previous studies in conformal symmetries have mainly used the $1+3$ decomposition of spacetime and the orthonomal tetrad. We follow this approach so that comparison with earlier results are made easier. Our intention is to excavate the most general class of conformal Killing vectors, that lies in the two-dimensional subspace described by the null and radial coordinates, that are admitted by the generalised Vaidya geometry. Subsequently we find the most general class of generalised Vaidya mass functions that give rise to such conformal symmetry. This analysis transparently brings out the reason for some well known subclasses of generalised Vaidya geometry, like pure Vaidya or charged Vaidya solutions, admitting  only homothetic Killing vectors but no proper conformal Killing vectors with non constant conformal factors. We also study the continual collapse of generalised Vaidya spacetimes that admit conformal symmetry and show that if the central singularity is naked or in the transition from a naked singularity to a covered one, the spacetime in the vicinity of the central singularity is almost self similar. \\

The paper is organised as follows: In Section \ref{one} we discuss the generalised Vaidya metric and the associated matter content. We generate the master equation governing the existence of the conformal symmetry in Section \ref{two} and show that a solution always exists. Special cases corresponding to particular mass functions are listed in Section \ref{three}. In section \ref{four} we study the collapse of generalised Vaidya spacetimes that admit proper conformal symmetry. Some general comments are given in Section \ref{five}. \\

Unless otherwise specified, we use natural units ($c=8\pi G=1$) throughout this paper, Latin indices run from 0 to 3. 
The symbol $\nabla$ represents the usual covariant derivative and $\partial$ corresponds to partial differentiation. 
We use the $(-,+,+,+)$ signature and the Riemann tensor is defined by
\begin{equation}
R^{a}{}_{bcd}=\Gamma^a{}_{bd,c}-\Gamma^a{}_{bc,d}+ \Gamma^e{}_{bd}\Gamma^a{}_{ce}-\Gamma^e{}_{bc}\Gamma^a{}_{de}\;,
\end{equation}
and the Ricci tensor is obtained by contracting the {\em first} and the {\em third} indices of the Riemann tensor.
  
\section{Generalised Vaidya Spacetimes \label{one}}

We know, the only types of physical matter fields are those whose energy momentum tensor either have one timelike and three spacelike eigenvectors or have double null eigenvectors \cite{HE}. The former  (which includes dust, perfect fluid form of matter) is called {\it Type I} matter field while the later (which includes radiation) is known as {\it Type II} matter field.
Most general spherically symmetric line element for arbitrary combination of {\it Type I} matter fields and {\it Type II} matter fields is given as \cite{israel}
\begin{eqnarray}
ds^2& = &-e^{2\psi (v,r)}\left[1-\frac{2m(v,r)}{r}\right ]dv^2+2\epsilon e^{\psi (v,r)}dvdr \nonumber\\
&&+r^2(d\theta^2+\sin^2\theta d\phi ^2), \:(\epsilon=\pm 1),\label{generalized-vaidya}
\end{eqnarray}
where $m(v,r)$ is the {\it Misner-Sharp} mass function that relates to the gravitational energy inside a given radius $r$ \cite{Lake1990}. When $\epsilon=+1$, the null coordinates $v$ is the Eddington advanced time, where $r$ is decreasing towards the future along a ray $v=Const$.  This depicts ingoing (or collapsing) null congruence with negative volume expansion. For $\epsilon=-1$,  the null coordinate $v$ is the Eddington retarded time and it depicts an outgoing null congruence with positive volume expansion. In this paper without any loss of generality we will consider $\epsilon=-1$ throughout.

Wang and Wu  \cite{Wang} established that there always exist classes of specific combinations of  {\it Type I}  and {\it Type II} matter fields, that makes the metric function $\psi (v,r)= 0$. This gives rise to the generalised Vaidya geometry with the line element
\begin{equation}
ds^2 = -\left(1-\frac{2m(v,r)}{r}\right )dv^2-2dvdr +r^2d\Omega^2. \label{line-element}
\end{equation}
This line element is the generalisation of well known Vaidya spacetime \cite{Vaidya}, and this generalisation comes from the observation that the energy-momentum tensor for these matter fields are linear in terms of the mass function. Therefore any  linear superposition of particular solutions is also a solution to the field equations. Hence, we can easily construct the monopole-de Sitter-charged Vaidya solutions and the Husain solutions that falls in this class. Generalised Vaidya spacetimes are also widely used in describing the formation of regular black holes \cite{sean} and black holes with closed apparent horizon \cite{frolov}.

For these spacetimes, the non-vanishing components of the Einstein tensor can be written as
\begin{subequations}
\begin{eqnarray}
G^v_v &=& G^r_r=-\frac{2m_{,r}}{r^2},\label{Einstein1}\\
G^r_v &=& \frac{2m_{,v}}{r^2},\label{Einstein2}\\
G^{\theta}_{\theta} &=& G^{\phi}_{\phi}=-\frac{m_{,rr}}{r}\;.\label{Einstein3}
\end{eqnarray}
\end{subequations}
We define  the null vectors  $l_{\mu}$ and $n_{\mu}$ in the following way
\begin{eqnarray}
l_{\mu} =\delta ^0_{\mu}, \quad  n_{\mu}=\frac{1}{2}\left [1-\frac{2m(v,r)}{r}\right]\delta ^0_{\mu}-\delta^1_{\mu}
\end{eqnarray}
where $l_{\mu}l^{\mu} =n_{\mu}n^{\mu}=0$ and $l_{\mu}n^{\mu}=-1$.
Using these null vectors we can now define the orthonormal basis
\begin{eqnarray}\label{orthonormal}
    E^{\mu}_{(0)} = \frac{l^{\mu} + n^{\mu}}{\sqrt{2}}, &&\quad E^{\mu}_{(1)} = \frac{l^{\mu} - n^{\mu}}{\sqrt{2}},\nonumber \\
     \quad E^{\mu}_{(2)} = \frac{1}{r}\delta^{\mu}_2, &&\quad E^{\mu}_{(3)} = \frac{1}{r\sin\theta}\delta^{\mu}_3,
\end{eqnarray}
Using the Einstein field equations, the corresponding energy momentum tensor, when projected to this orthonormal basis can be written in the form 
\begin{equation}\label{EMTMatrix}
    \left[T_{(\mu)(\nu)}\right] = \left[
                                \begin{array}{cccc}
                                  \frac{\mu}{2}+ \rho & \frac{\mu}{2} & 0 & 0 \\
                                  \frac{\mu}{2} & \frac{\mu}{2} - \rho & 0 & 0 \\
                                  0 & 0 & P& 0 \\
                                  0 & 0 & 0 &P \\
                                \end{array}
                              \right].
\end{equation}
where 
\begin{equation}
\mu=\frac{2m_{,v}}{r^{2}},\ \ \  \rho=\frac{2m_{,r}}{r^{2}}, \ \ \  P =\frac{2m_{,rr}}{r^{2}}
\end{equation}
This is the form of energy momentum of a specific combination of {\it Type I} and{\it Type II} fluid as defined in \cite{HE}, with the following energy conditions:
\begin{enumerate}
  \item \emph{The weak and strong energy conditions}:
  \begin{equation}\label{strongweakenergy}
   \mu\geq 0, \quad \rho\geq 0, \quad P \geq 0, \quad (\mu\neq 0).
  \end{equation}
  \item \emph{The dominant energy conditions}:
  \begin{equation}\label{dominantenergy}
    \mu\geq 0, \quad \rho \geq P \geq 0, \quad (\mu\neq 0).
  \end{equation}
\end{enumerate}
We can suitably choose the mass function $m(v,r)$, to satisfy all these energy condition. For Vaidya spacetimes, when
$m = m(v)$, the fluid is a pure {\it Type II} fluid, and the energy conditions are satisfied when $\mu \geq 0$. On the other hand when $m = m(r)$ we have $\mu = 0$, and the matter field degenerates to a pure {\it Type I} fluid with the usual energy conditions. 

\section{Existence of Conformal Killing Vectors}\label{two}

Any spacetime (with coordinates $x^a$ and metric $g_{ab}$) is said to possess a {\it conformal Killing vector} (CKV) `$\mathbb{X}$', if it solves the following conformal Killing equation
\be
L_{\mathbb{X}}g_{ab}=S(x^a)g_{ab}\,. \label{CKE}
\ee
If the components of the vector $\mathbb{X}$ is represented by $X^a$ then the above equation simplifies to the set of equations
\be
\nabla_{(a}X_{b)}=Sg_{ab}\,.\label{CKE1}
\ee
We know the CKV becomes a Killing vector when $S=0$. In that case the metric remains invariant when it is Lie dragged along this vector field.  Homothetic Killing vector is a special case of conformal Killing vector if the conformal function $S$ is a non-zero constant. Then without any loss of generality we can always take $S$ to be unity. The spacetimes that possess a homothetic Killing vector are called {\it self similar spacetimes}. Henceforth we will denote a CKV as {\it proper} if $S$ is a non-constant function of the coordinates.

We would now like to find the existence of non-trivial CKV's in the $(v,r)$ subspace of a generalised Vaidya spacetime. We note that since the spacetime is spherically symmetric, the $(\theta,\phi)$ subspace will have the usual symmetries of a 2-sphere. Therefore any non-trivial symmetry must lie in the $(v,r)$ subspace. Thus, we look for a CKV of the form 
\be
\mathbb{X}=A(r,v)\partial_v+B(r,v)\partial_r\,, \label{CKV1}
\ee
where $A(r,v)$ and $B(r,v)$ are unknown functions to be determined by solving the conformal Killing equations (\ref{CKE1}). We note that due to the spherical symmetry and the form of the CKV chosen, (\ref{CKE1}) becomes a set of four non-trivial equations. The ($\theta,\theta$) component gives
\be
B=Sr\,, \label{CKE2}
\ee
which is just a definition of the conformal factor $S$ in terms of the vector component. The ($r,r$) component gives
\be
A_{,r}=0\,,  \label{CKE3}
\ee
which constraints the function $A\equiv A(v)$. We note that for proper CKV and homothetic Killing vectors, we may use the scaling freedom of the null coordinate `$v$' to write $A(v)=v$, without any loss of generality. (Although this is not true for Killing vectors where $A(v)$ is constant, we exclude that singular case here).  Now using the above two equations the ($r,v$) component becomes
\be
1+B{_r}-2\frac{B}{r}=0 \,.  \label{CKE4}
\ee
Solving the above equation for the function $B(r,v)$, we get 
\be\label{sol1}
B(r,v)=g(v)r^2+r\,,
\ee
where $g(v)$ is an arbitrary function of integration. This completes the demonstration of the following proposition:
\begin{prop}
The most general class of proper CKV in the $(v,r)$ subspace of a generalised Vaidya spacetime is of the form $\mathbb{X}=v\partial_v+(gr^2+r)\partial_r$. The corresponding conformal factor is given by $S=1+rg$.
\end{prop}
From the above proposition, it is obvious that when $g=0$, the CKV becomes homothetic. We would now like to find out what classes of mass function can admit these general CKV's in $(v,r)$ sunspace. We use the ($v,v$) component of (\ref{CKE1}) and using (\ref{CKE2}, \ref{CKE3}  and \ref{CKE4}), to obtain
\be
\left(1-\frac{2m}{r} \right)B_{,r} + B_{,v} + \frac{B}{r}\left( m_{,r} + \frac{m}{r} - 1 \right) + \frac{m_{,v}v}{r} = 0 \,.  \label{CKE5}
\ee
Now the interesting point about the above equation is: given any function $B(r,v)$, the equation becomes a quasilinear equation for the function $m(r,v)$, whose solution is guaranteed. 
Therefore we plug in the form of $B(r,v)$ from equation (\ref{sol1}) to get the required quasilinear equation 
\be\label{m}
vm_{,v}+\left(gr^2+r\right)m_{,r}-m\left(3gr+1\right)+\left(gr^2+g_{,v}r^3\right)=0
\ee
The general solution of the above completes the demonstration of the following:
\begin{prop}
The most general class of Misner-Sharp mass function that allows for a proper CKV in the $(v,r)$ subspace of a generalised Vaidya spacetimes is given as
\be\label{m2}
m(r,v)=rx^2\left(F\left(\int gdv+x\right)+\int x\left[g_{,v}+g\right]dv\right)\
\ee
where $x\equiv v/r$ and the function $F$ is an arbitrary function of integration.
\end{prop}

We can immediately state a corollary to the above proposition:
\begin{cor}
If pure Vaidya spacetimes with $m(r,v)=m(v)$ possess a CKV, we must have $g=0$. In this case the CKV is a homothetic Killing vector. Thus although pure Vaidya spacetimes allows for self similar solutions, they do not admit a proper conformal symmetry.
\end{cor}
 
 The above corollary is consistent with the result by Lewandowski \cite{lewandowski1990conformal}, that a spacetime with pure null radiation cannot contain a proper CKV. Note that the absence of proper CKV's will also apply to simple extensions of the Vaidya spacetimes considered by Wang and Wu \cite{Generalizedvaidya1999}, including charged Vaidya and Husain metrics. 

\section{Examples of some special cases}\label{three}

In this section we will look into some physically interesting special cases for which we can exactly integrate the equation (\ref{m2}) to get some specific class of mass functions. Again note that by rescaling $A(v)\equiv v$, we have excluded the singular case of Killing vectors. 
 
 \subsection{Case of Homothety}
 
 Homothetic Killing vector is a special case of conformal Killing vector if the conformal function $S$ is a non-zero constant. Without any loss of generality we can always take $S$ to be unity. The spacetimes that posses a homothetic Killing vector is called {\it self similar spacetimes}. Now from equation  (\ref{CKE2}) we get $B=r$. So then the homothetic Killing vector is of the form

\be
 \mathbb{X}=v\partial_v + r\partial_r\,,
 \ee 
 and equation (\ref{CKE5}) reduces to
 \be
 m_{,r}+\frac{m_{,v}v}{r}-\frac{m}{r}=0\,. \label{homo}
 \ee
 The general solution of the above equation is
 \be
 m(v,r)=rF(x)\,,
 \ee
 where $x=v/r$ and $F$ is an arbitrary function. Thus the only self similar Vaidya solution is for $m(v)=\lambda v$ for some constant $\lambda$.

 Similarly for charged Vaidya the self similar spacetime is obtained when
 \be 
 m(v,r)=\lambda_1 v + \lambda_2 \frac{v^2}{r}. 
 \ee
 In general, any mass function of the form 
 \be
 m(r,v)=\sum_{n=-\infty}^{\infty} \frac{m_{n}(v)}{r^{n-1}} 
 \ee
 will allow for a homothety iff
 \be
 m_{n}(v) = \lambda_n v^{n} . 
 \ee

 \subsection{Case of $S=S(r)$} 
 
Next we look for spacetimes with proper conformal symmetry, where the conformal factor is a function of the radial coordinate only. Thus substituting $g(v)=K$ in equation (\ref{m}) and solving, we get the general class of solutions
 \be
 m(r,v)= G\left(x(Kr+1)\right)\left[K^2r^3+2Kr^2+r\right]+\frac12 r\,
 \ee
where $x\equiv v/r$ and the function $G$ is an arbitrary function of integration. It is interesting to see that a dynamic spacetime of the above form admits a static conformal factor. However when $G=const.$, we can always have a subclass of anisotropic deSitter spacetime of the form 
\be
m(r,v)=m(r)=K^2r^3+2Kr^2+\frac32 r\,,
\ee
which admits proper CKV with static conformal factor. 

 \subsection{Case of $S=1+v^nr^2$} 
 
 Here we take the function $g(v)=v^n$. Again substituting in equation (\ref{m}) and solving, we get the general class of solutions

\ba
m \left( r,v \right)&=&-\frac {{r}^{2}}{ \left( n+1 \right) ^{3}{v}^{3
}} \left( -vr{\it H} \left( \Phi \right) + \frac12 {v}^{2\,n+3}r(n+1)\right.\nonumber\\
&&\left. +{v}^{n+2} \left( n+1 \right) ^2 \left( nr+v \right) \right) \,
\ea
where we have 
\be
\Phi= \left( {\frac {v \left( {v}^{n}r+n+1 \right) }{r \left( n+1 \right) }} \right) \,,
\ee
and $H$ is again an arbitrary function of integration. 

\section{Collapse of generalised Vaidya spacetimes admitting proper conformal symmetry}\label{four}

In \cite{Maombi2}, the continual gravitational collapse of generalised Vaidya spacetimes was investigated in detail in the context of Cosmic Censorship Conjecture. It was shown that all the non-central singularities must be covered behind the apparent horizon. However the central singularity at ($v=0,r=0$) can be naked, subjected to certain conditions on the mass function. These conditions are as follows:
\begin{enumerate}
\item $m_0 \equiv \lim\limits_{v\to 0, r \to 0}m(v,r)=0$.
\item $\dot{m}_0\equiv \lim\limits_{v\to 0, r \to 0}m_{,v}$ and  $m'_0 = \lim\limits_{v\to 0, r \to 0}m_{,r}$ are well defined.
\item The quantities $b_{\pm} \equiv \frac{(1-2m'_0)\pm\sqrt{(1-2m'_0)^2 - 16\dot{m}_0}}{4\dot{m}_0}$ must be positive and real.
\end{enumerate}
Now we would like to investigate the the final outcome of the continual collapse of those generalised Vaidya spacetimes that admit a proper conformal symmetry, with a class of mass functions given by (\ref{m2}). By the arbitrariness of the functions $F$ and $g$, it is clear that once the first two conditions are satisfied there always exist sets of non-zero measure in the functional space for which the third condition is also satisfied, and hence  for these spacetimes Cosmic Censorship Conjecture may be violated. However the first two conditions give an interesting insight. They specifically force the functions $F$ and $g$ to be at least $C^1$ functions in the limit of the central singularity ($v=0,r=0$). Therefore $g(0)$ must be well defined and bounded. By Proposition 1, we can now immediately see that for small enough values of the coordinate $r$, (where we can neglect the $O(r^2)$ term), the proper CKV becomes a HKV as $\mathbb{X}\sim v\partial_v + r\partial_r$. This observation enables us to state the following proposition:
\begin{prop}
Let the end state of a continual gravitational collapse of generalised Vaidya spacetimes that admit a proper conformal symmetry, is a naked singularity or in the transition region of naked singularity and black hole phases. In that case there exist an open set about the central singularity where the spacetime is almost self similar. Also the almost homothetic Killing vector becomes null at the central singularity, indicating the the formation of a horizon at the centre. 
\end{prop}
In the above context there is a remarkable similarity between generalised Vaidya spacetimes, Lake's results \cite{Lake} and Choptuik's numerical results of massless scalar field collapse \cite{Choptuik}.

\section{Discussion}\label{five}

In this paper, we performed a detailed investigation on the conformal symmetry in the ($v,r$) subspace of generalised Vaidya spacetimes. Due to the spherically symmetry, we noted that the $(\theta,\phi)$ subspace will have the usual symmetries of a 2-sphere and hence any non-trivial symmetry must lie in the $(v,r)$ subspace. We integrated the conformal Killing equations to explicitly find the most general form of CKV that may lie in the $(v,r)$ subspace and subsequently found the most general class of mass functions that admit such a symmetry. \\

From our analysis it is very transparent that some very well known solutions of the generalised Vaidya class (for example pure Vaidya or charged Vaidya), although admitting homothetic symmetry, do not admit any proper conformal symmetry. We also regained the mass function for the well known class of anisotropic deSitter spacetimes that admit a proper CKV. And finally we studied the continual gravitational collapse of generalised Vaidya spacetimes that admit a proper conformal symmetry, to show that there exist set of non-empty measure in the parameter space of the mass function for which Cosmic Censorship Conjecture is violated. However when such scenario happens, there exist an open set about the central singularity where the spacetime is almost self similar and the almost homothetic Killing vector becomes null at the central singularity as the horizon forms. This result is at par with the observations of Lake and Choptuik. \\

A study of the complete set of conformal Killing vectors in generalised Vaidya spacetimes will definitely shed more light on the above results and we keep that for future works. 

\section {Acknowledgement \label{acknowledge}}
The authors would like to thank Dr. AMK Nzioki for GRTensor help.
We are indebted to the National Research Foundation and the University of KwaZulu-Natal for financial support.
SDM acknowledges that this work is based upon research supported by the South African Research Chair Initiative of the Department of Science and Technology.

 \end{document}